\documentclass[aps,prl,reprint,amsmath,amssymb,showpacs,superscriptaddress]{revtex4-1}
\usepackage{CJK}
\usepackage{graphicx}% Include figure files
\usepackage{dcolumn}% Align table columns on decimal point
\usepackage{bm}% bold math
\usepackage{mathrsfs}   % added by S.B. Wang to use \mathscr{L}_{N\pi}
\usepackage{color,xcolor}  % added by S.B. Wang to use \color{red}
\usepackage{multirow}  % added by S.B. Wang to use multirow in table
\usepackage{booktabs} % added by S.B. Wang to use \cmidrule(r){5-6}

\usepackage{hyperref}% add hypertext capabilities
\newcommand{\la}{\langle}
\newcommand{\ra}{\rangle}

\newcommand{\bbm}{\begin{bmatrix}}
\newcommand{\ebm}{\end{bmatrix}}
\newcommand{\bBm}{\begin{Bmatrix}}
\newcommand{\eBm}{\end{Bmatrix}}
\newcommand{\bpm}{\begin{pmatrix}}
\newcommand{\epm}{\end{pmatrix}}

\begin{document}
% \begin{CJK*}{UTF8}{song} % Use default fonts from CJK (see below)


\title{Asymmetric Nuclear Matter and Neutron Star Properties in Relativistic \emph{ab initio} Theory in the Full Dirac Space}

\author{Sibo Wang}
\affiliation{Department of Physics, Chongqing University, Chongqing 401331, China}

\author{Hui Tong}
\affiliation{College of Physics and Materials Science, Tianjin Normal University, Tianjin 300387, China}
\affiliation{Strangeness Nuclear Physics Laboratory, RIKEN Nishina Center, Wako, 351-0198, Japan}

\author{Qiang Zhao}
\affiliation{Center for Exotic Nuclear Studies, Institute for Basic Science, Daejeon 34126, Korea}

\author{Chencan Wang}
\affiliation{School of Physics, Nankai University, Tianjin 300071, China}

\author{Peter Ring}
\affiliation{Department of Physics, Technische Universit\"{a}t M\"{u}nchen, D-85747 Garching, Germany}

\author{Jie Meng}
\email{mengj@pku.edu.cn}
\affiliation{State Key Laboratory of Nuclear Physics and Technology, School of Physics, \\
Peking University, Beijing 100871, China}
\affiliation{Yukawa Institute for Theoretical Physics, Kyoto University, Kyoto 606-8502, Japan}

\date{\today}

\begin{abstract}
The long-standing controversy about the isospin dependence of the effective Dirac mass in \emph{ab} \emph{initio} calculations of asymmetric nuclear matter is clarified by solving the relativistic Brueckner-Hartree-Fock equations in the full Dirac space. The symmetry energy and its slope parameter at the saturation density are $E_{\text{sym}}(\rho_0)=33.1$\ MeV and $L=65.2$ MeV, in agreement with empirical and experimental values.
Further applications predict the neutron star radius $R_{1.4M_\odot}\approx 12$\ km and the maximum mass of a neutron star  $M_{\text{max}}\leq 2.4M_\odot$.

\end{abstract}

%\pacs{26.60.-c,21.65.Ef,21.60.Jz,21.60.De}

%21.60.Jz Nuclear Density Functional Theory and extensions
%21.10.Gv Nucleon distributions and halo features
%21.30.-x Nuclear forces
%21.10.-k Properties of nuclei; nuclear energy levels
%21.10.Re Collective levels
%21.60.Ev Collective models
%21.60.Cs Shell model
%21.45.Ff Three-nucleon forces
%23.20.-g Electromagnetic transitions
%23.20.Js Multipole matrix element
%26.60.-c Nuclear matter aspects of neutron stars
%21.65.Ef Symmetry energy
%21.60.-n Nuclear structure models and methods
%27.20.+n  6 A 19
%27.60.+j 90  A 149
%27.50.+e 59  A  89
%21.60.De Ab initio methods
%\keywords{}

\maketitle
%\end{CJK*}


%=======================================================================================
%\section{Introduction}\label{SecI}
%=======================================================================================

% Asymmetric nuclear matter is important

\emph{Introduction.} Neutron stars provide a unique and natural laboratory for dense nuclear matter at extreme conditions that cannot be reproduced in any terrestrial laboratory~\cite{Shapiro1983,Lattimer2004_Science304-536}.
The observation of gravitational waves from a binary neutron star merger~\cite{Abbott2017_PRL119-161101,Abbott2018_PRL121-161101,Abbott2020_ApJL892-L3} and the combinations of mass and radius measurements of neutron stars~\cite{Lattimer2012_ARNPS62-485,Ozel2016_ARAA54-401} have placed substantial constraints on the behavior of cold nuclear matter at suprasaturation density~\cite{Annala2018_PRL120-172703}.
Together with the information of nuclear structure experiments~\cite{Tsang2012_PRC86-015803,Baran2005_PR410-335} and heavy-ion collisions (HICs)~\cite{Danielewicz2002_Science298-1592,LI-BA2008_PR464-113}, %{\color{red}probing the equation of state (EOS) of nuclear matter at lower densities,} 
astrophysical observations are constantly revealing the mysterious properties of dense matter~\cite{Fattoyev2018_PRL120-172702,Soumi2018_PRL121-091102,LI-Baoan2021_Universe7-182}.

Asymmetric nuclear matter (ANM) has attracted considerable attention since its equation of state (EOS) and the density dependence of the symmetry energy provide important microscopic inputs for the investigation of the structure of neutron star interiors~\cite{Lattimer2004_Science304-536,LATTIMER2007_PR442-109}, neutron star mergers~\cite{Baiotti2019_PPNP109-103714}, and the dynamics of supernova explosions~\cite{Oertel2017_RMP89-015007}.

On the theoretical side, on the basis of nuclear density functional theories (DFTs)~\cite{Brown2000_PRL85-5296,RocaMaza2011_PRL106-252501,ZHANG-Z2015_PRC92-031301, Zhao2016_PRC94-041302,TONG-H2020_PRC101-035802}, important correlations have been found between astrophysical as well as nuclear quantities and the properties of ANM. They start from phenomenological density-dependent effective nucleon-nucleon ($NN$) interactions in the medium.
These effective interactions are often determined by fitting to the ground-state properties of finite nuclei and the saturation properties of symmetric nuclear matter (SNM).
This leads to the fact that they are not well constrained in the extreme conditions of high density or large isospin asymmetry~\cite{CHEN-LW2015_EPJW88-00017}. In this situation, \emph{ab initio} calculations based on realistic $NN$ interactions are expected to give better predictions.

% RBHF theory in the full Dirac space

Relativistic Brueckner-Hartree-Fock (RBHF) theory is one of the most successful \emph{ab initio} theories based on bare two-body forces only~\cite{Shen-SH2016_CPL33-102103,SHEN-SH2019_PPNP109-103713}. In the relativistic framework, it contains the important $Z$ diagram~\cite{Brown1987_CNPP17-39}, an effective three-body force generated by a virtual nucleon-antinucleon excitation.

Since the pioneering work of the Brooklyn group~\cite{Anastasio1980_PRL45-2096,Anastasio1981_PRC23-2273}, RBHF calculations are primarily performed with positive-energy states (PESs), because the construction of $NN$ interaction matrix elements in full Dirac space, i.e., between negative-energy states (NESs) and PESs, and  the corresponding solution of the in-medium scattering equation are rather complicated.
To compensate for the incompleteness of the Dirac space, different approximations have been introduced to extract the effective single-particle potentials~\cite{Brockmann1990_PRC42-1965,Gross-Boelting1999_NPA648-105,Schiller2001_EPJA11-15} necessary for the self-consistent solution of the Hartree-Fock equation.
% These approximations have been applied to solve for symmetric~\cite{Brockmann1990_PRC42-1965,LI-GQ1992_PRC45-2782,Gross-Boelting1999_NPA648-105,Schiller2001_EPJA11-15,WANG-CC2020_JPG47-105108} and asymmetric nuclear matter~\cite{Engvik1994_PRL73-2650,MA-ZY2002_PRC66-024321,Alonso2003_PRC67-054301,VanDalen2004_NPA744-227,
% Sammarruca2005_PRC71-064306,VanDalen2007_EPJA31-29,Katayama2013_PRC88-035805,TONG-H2018_PRC98-054302}.
However, it turned out that they cannot uniquely determine the single-particle properties~\cite{Nuppenau1989_NPA504-839}. 
Contradictory results for the isospin dependence of the Dirac mass are found between two frequently used approximations~\cite{Ulrych1997_PRC56-1788}.
The momentum-independence approximation predicts the proton-neutron Dirac mass splitting in isospin asymmetric matter is $M_{D,n}^* > M_{D,p}^*$, while the projection method leads to the opposite sign $M_{D,n}^* < M_{D,p}^*$.
% Contradictory results for the isospin dependence of single-particle potentials are found between two frequently used approximations~\cite{Ulrych1997_PRC56-1788}.
Therefore, to clarify the properties of asymmetric nuclear matter, it is necessary to solve the RBHF equations in the full Dirac space~\cite{Poschenrieder1988_PRC38-471,Huber1995_PRC51-1790,deJong1998_PRC58-890}.

Recently, a self-consistent RBHF calculation in the full Dirac space has been achieved for SNM~\cite{WANG-SB2021_PRC103-054319}. It avoids the approximations applied in the RBHF calculations in the Dirac space with PESs only. The saturation properties of SNM found in this way are in good agreement with the empirical values. 
In this Letter, we develop the RBHF theory in the full Dirac space for ANM and present the results of ANM and their consequences for the mass-radius relations of neutron stars.

%=======================================================================================
%\section{Theoretical framework} \label{SecII}
%=======================================================================================

\emph{Theoretical framework.} In the RBHF theory, the nucleon inside the nuclear medium is viewed as a dressed particle due to its two-body interaction with the surrounding nucleons.
The single-particle motion of a nucleon with rest mass $M$, momentum $\bm{p}$, and single-particle energy $E_{\bm{p}}$ is depicted by the Dirac equation.
\begin{equation}\label{eq:DiracEquation}
  \left[ \bm{\alpha}\cdot\bm{p}+\beta \left(M+\Sigma(\bm{p})\right) \right] \psi(\bm{p})
  = E_{\bm{p}}\psi(\bm{p}),
\end{equation}
where $\Sigma$ is the single-particle potential (the self-energy) in the Dirac space. For simplicity, spin and isospin indices are neglected. 
The Dirac spinors are denoted by $\psi(\bm{p})$.
For each $\bm{p}$ there are a solution with positive energy (PES) and one with negative energy (NES).

In the RBHF scheme, the single-particle potential operator $\Sigma$ in Dirac space is calculated as an integral over the effective interaction, the $G$ matrix,

\begin{equation}\label{eq:Sigma}
\Sigma(p) = \ \int^{p_F}_0 \frac{d^3p'}{(2\pi)^3}
    \la \bar{\psi}(\bm{p}) \bar{\psi}(\bm{p}')| \bar{G}(W)|\psi(\bm{p})\psi(\bm{p}')\ra.
\end{equation}
Here, the integral runs over all occupied states in the Fermi sea ($|\bm{p}'|\le p_F$) and, for simplicity, spin and isospin indices are neglected. 
The starting energy is denoted by $W$. 
For further details, see Ref.~\cite{WANG-SB2021_PRC103-054319}.

The effective $NN$ interaction $G$ is the basic ingredient of RBHF theory.
In the nonrelativistic Brueckner-Hartree-Fock theory, it is an effective scattering matrix in the nuclear medium, found as the solution of the Bethe-Goldstone equation~\cite{Bethe1957_PRSA238-551,Brueckner1958_PR109-1023}.
Here, the Pauli operator $Q$ excludes, in the intermediate states, scattering processes to occupied states below the Fermi surface. 

In relativistic scattering processes, the scattering matrix is determined by the four-dimensional Bethe-Salpeter equation~\cite{Salpeter1951_PR84-1232}.
There are several three-dimensional reductions to this equation~\cite{Blankenbecler1966_PR142-1051,Thompson1970_PRD1-110}.
Nowadays, in most of the applications of the RBHF theory, the $G$ matrix is obtained by solving the in-medium covariant Thompson equation~\cite{Brockmann1990_PRC42-1965},
\begin{widetext}
\begin{equation}\label{eq:ThomEqu}
 G(\bm{q}',\bm{q}|\bm{P},W)
  =\ V(\bm{q}',\bm{q}|\bm{P}) +
  \int \frac{d^3k}{(2\pi)^3}  V(\bm{q}',\bm{k}|\bm{P})
\frac{Q(\bm{k},\bm{P})}{W-E_{\bm{P}+\bm{k}}-E_{\bm{P}-\bm{k}}} G(\bm{k},\bm{q}|\bm{P},W).
\end{equation}
\end{widetext}
Here, $\bm{P}=\frac{1}{2}({\bm k}_1+{\bm k}_2)$ and $\bm{k}=\frac{1}{2}({\bm k}_1-{\bm k}_2)$ are the center-of-mass and the relative momenta of the two interacting nucleons with the momenta ${\bm k}_1$ and ${\bm k}_2$.
The initial, intermediate, and final relative momenta of the two nucleons are $\bm{q}, \bm{k}$, and $\bm{q}'$, respectively. 
The starting energy is denoted by $W$ and the $NN$ scattering in the nuclear medium is restricted by the Pauli operator $Q(\bm{k},\bm{P})$.

Equations \eqref{eq:DiracEquation}--\eqref{eq:ThomEqu} constitute a coupled system that has to be solved in a self-consistent way.
After the convergence is satisfied, the binding energy per nucleon for ANM can be calculated straightforwardly~\cite{VanDalen2004_NPA744-227,Katayama2013_PRC88-035805,TONG-H2018_PRC98-054302}.

In previous RBHF calculations~\cite{Brockmann1990_PRC42-1965,Gross-Boelting1999_NPA648-105,Schiller2001_EPJA11-15}
the Thompson equation~\eqref{eq:ThomEqu} is solved in the Dirac space with PES only.
Because relativistic scattering algorithms are relatively complicated, the starting point for such calculations was the free scattering algorithm used for the derivation of a relativistic $NN$ potential from the experimental phase shifts~\cite{Erkelenz1971_NPA176-413,Erkelenz1974_PR13-191}. By obvious reasons this algorithm is restricted to the scattering of particles with positive-energy and negative-energy solutions (scattering of antiparticles) are not considered. The results of such calculations are the matrix elements of the scattering matrix ($T$ matrix) and the corresponding phase shifts for proton and neutrons, i.e., for particles with positive energy. 
For the self-consistent solution of the RBHF equations, the situation is much more complicated. 
In each step of the iteration, we need, for the evaluation of the Dirac spinors in the medium by the solution of the Dirac equation \eqref{eq:DiracEquation}, not only the matrix elements of the potential $\Sigma$ for PESs $\Sigma^{++}$, but also matrix elements $\Sigma^{+-}$ between PESs and NESs and the elements between NESs $\Sigma^{--}$. This requires, in principle, a solution of the Thompson equation~\eqref{eq:ThomEqu} in the full Dirac space. 
Previous RBHF calculations avoid the calculation of $\Sigma^{+-}$ and $\Sigma^{--}$, and use several approximations~\cite{Brockmann1990_PRC42-1965,Gross-Boelting1999_NPA648-105,Schiller2001_EPJA11-15}. 
%for the self-consistent solution of Equations \eqref{eq:DiracEquation}, \eqref{eq:Sigma}, and \eqref{eq:ThomEqu}~

%=======================================================================================
%\section{Results and discussion}\label{SecIV}
%=======================================================================================

% fig 1: Dirac effective mass for neutron and proton
\begin{figure*}[htbp]
  \centering
  \includegraphics[width=14.0cm]{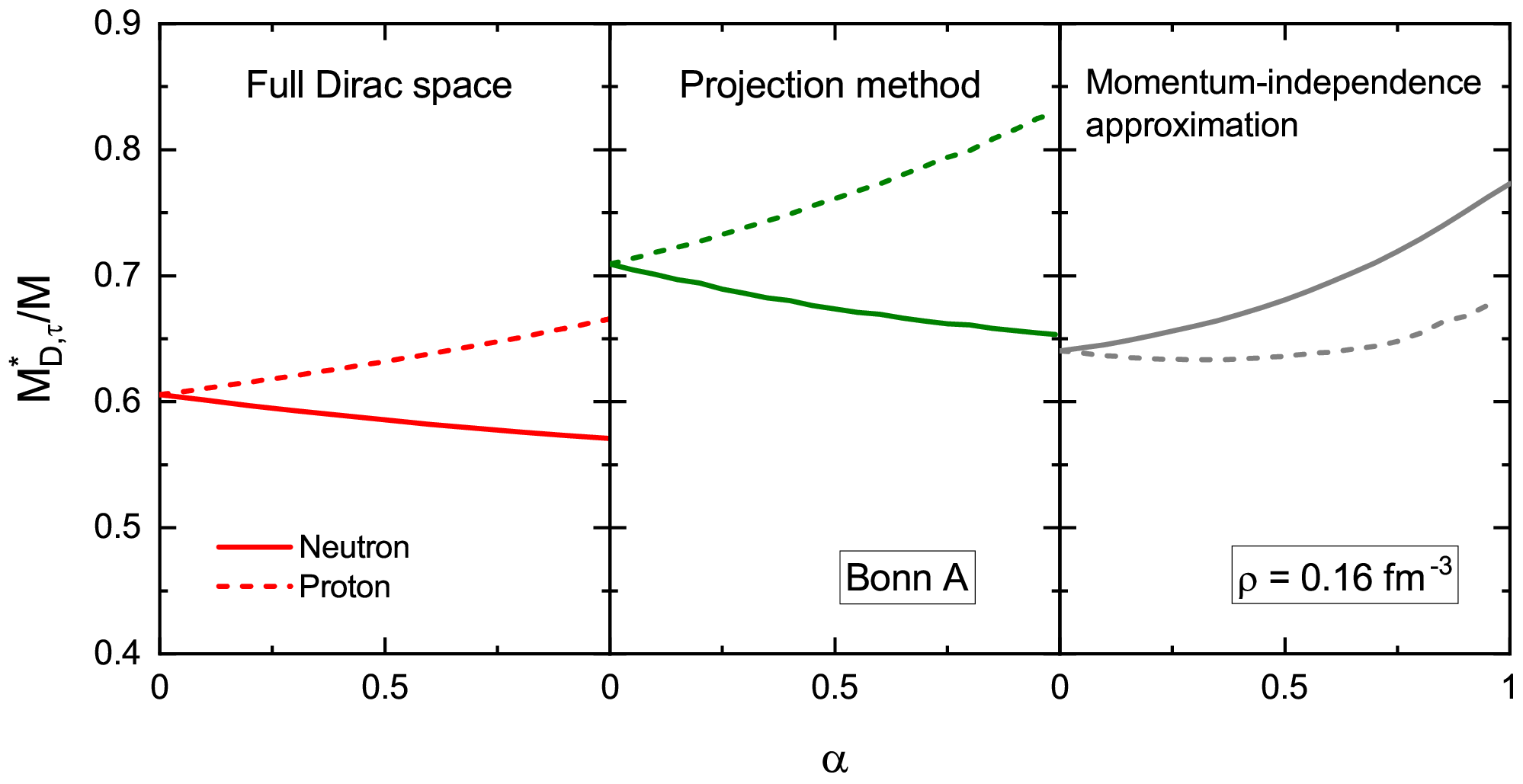}
  \caption{(Color online) Effective Dirac mass for the neutron (solid lines) and the proton (dashed lines) as functions of the asymmetry parameter $\alpha$ at $\rho=0.16\ \text{fm}^{-3}$ calculated by the RBHF theory in the full Dirac space (left), in comparison with the results obtained by RBHF calculations with PESs only using the projection method~\cite{Gross-Boelting1999_NPA648-105} (middle) and the momentum-independence approximation~\cite{Brockmann1990_PRC42-1965} (right). The Bonn-A potential~\cite{Machleidt1989_ANP19-189} is used.}
  \label{Fig1}
\end{figure*}
\emph{Results and discussion.} The essential feature of the RBHF theory in the full Dirac space is the fact that the properties of single-particle potential 
\begin{equation}\label{eq:SPP}
  \Sigma(\bm{p}) = U_{S}(p)+ \gamma^0U_{0}(p) + \bm{\gamma\cdot\hat{p}}\,U_{V}(p)
\end{equation}
can be determined uniquely. Here, $\hat{\bm{p}}=\bm{p}/p$ is the unit vector parallel to the momentum $\bm{p}$. 
The quantities $U_S(p)$, $U_0(p)$, and $U_V(p)$ are the scalar potential, the timelike part and the spacelike part of the vector potential.

The isospin dependence of the single-particle potential can be well illustrated by the relativistic effective Dirac mass, which is defined through the scalar part of the nucleon self-energy in the Dirac equation by $M_{D,\tau}^* = M + U_{S,\tau}$. 
The Dirac mass should not be confused with the definition of the nonrelativistic effective mass, which parametrizes the momentum and energy dependence of the single-particle potential~\cite{Jaminon1989_PRC40-354}.
% To illustrate the isospin dependence of the single-particle potential,
In the left panel of Fig.~\ref{Fig1}, the Dirac mass of the nucleon at the Fermi surface for protons and neutrons ($\tau=p,n$) obtained with the RBHF theory in the full Dirac space are plotted as functions of the asymmetry parameter $\alpha=(\rho_n-\rho_p)/\rho$ at the density $\rho=0.16\ \text{fm}^{-3}$.
It is found that, with the increasing of the asymmetry parameter, the Dirac mass for the neutron is decreasing, while for the proton, an opposite tendency is obtained.
As a result, $M^*_{D,n}<M^*_{D,p}$ with the isovector effective mass $(M^*_{D,p}-M^*_{D,n})/M = 0.095$ in pure neutron matter (PNM) is predicted in the full Dirac space.

The other two panels of Fig.~\ref{Fig1} contain approximations used in the literature, where the Thompson equation~\eqref{eq:ThomEqu} is solved only for PESs, and the potentials are determined approximately: The middle panel is obtained by the projection method~\cite{Gross-Boelting1999_NPA648-105} with the \emph{ps} representation for the subtracted $T$ matrix described in detail in Ref.~\cite{VanDalen2004_NPA744-227}.
It is noticed that, in comparison to the results in the full Dirac space, the projection method leads to a qualitatively consistent isospin dependence of the Dirac mass, but the amplitudes of $M^*_{D,n}$ and $M^*_{D,p}$ are overestimated.
The right panel of Fig.~\ref{Fig1} shows the results obtained with the momentum-independence approximation~\cite{Brockmann1990_PRC42-1965}, where the single-particle potentials are assumed to be independent of the momentum, and the spacelike part of the vector potential is neglected.
With this approximation, the scalar potential and the timelike part of the vector potential are extracted directly from the single-particle potential energies at two casually selected momenta, $0.7k^\tau_F$ and $k^\tau_F$.
It can be seen that $M^*_{D,n}>M^*_{D,p}$ is obtained for the entire region of the asymmetry parameter, which is contradictory to that calculated with the RBHF theory in the full Dirac space and the projection method.
As pointed out in Ref.~\cite{Ulrych1997_PRC56-1788}, a wrong sign for the isovector dependence of single-particle potentials is obtained by applying the momentum-independence approximation to asymmetric nuclear systems.
In short, by performing the RBHF calculation in the full Dirac space, the long-standing controversy of the isospin dependence of the effective Dirac mass, i.e., the opposite tendency predicted with the momentum-independence approximation and projection method, has been clarified. In addition, it is shown that the solution in the full Dirac space leads to a weaker isospin dependence of the Dirac mass splitting than the projection method.

% fig 2: symmetry energy

\begin{figure}[htbp]
  \centering
  \includegraphics[width=8.0cm]{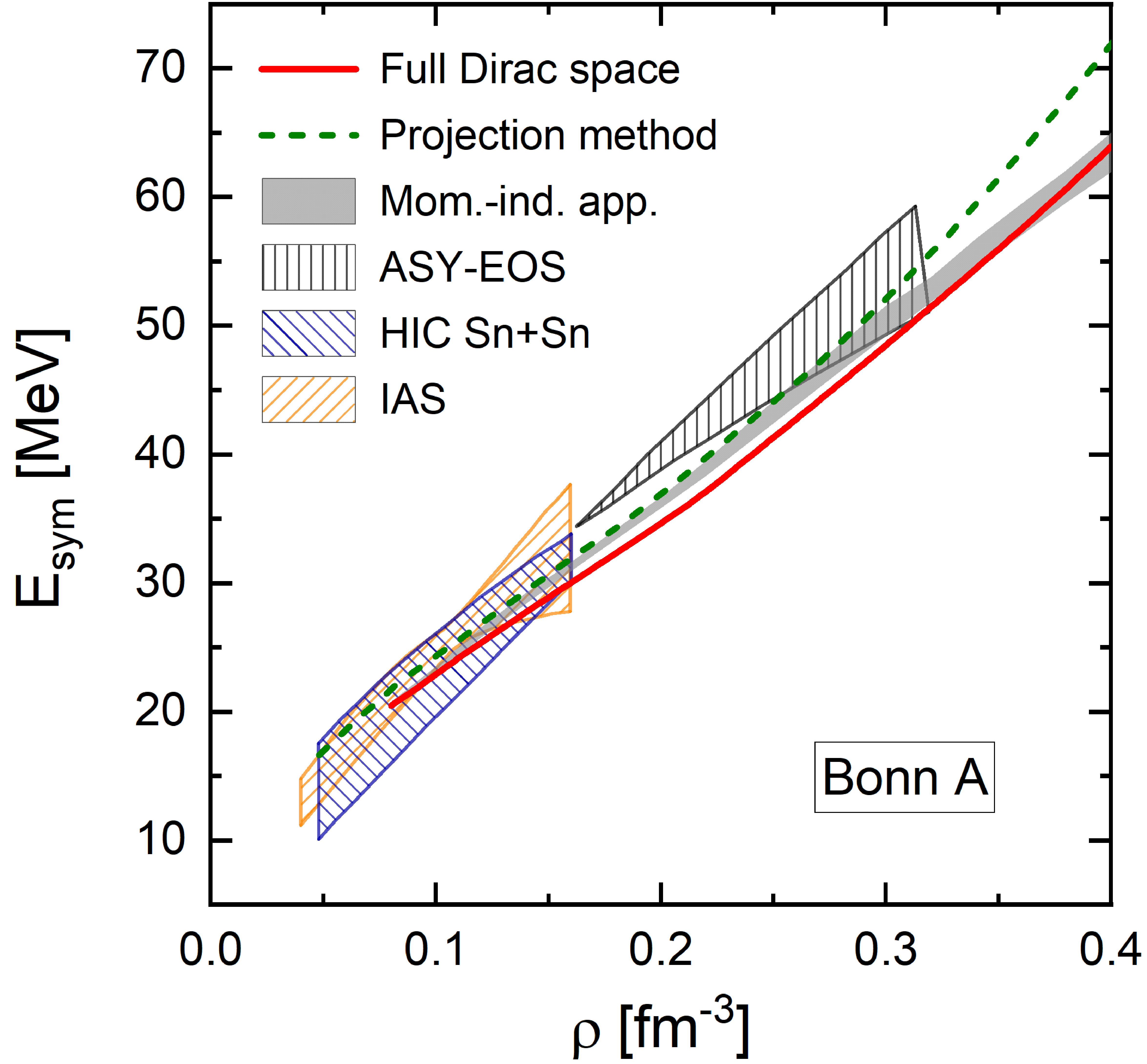}
  \caption{(Color online) The symmetry energy $E_{\text{sym}}$ as a function of the density $\rho$ calculated by the RBHF theory in the full Dirac space (red solid line), in comparison with the results obtained by the RBHF calculation with the projection method (olive dashed line) and the momentum-independence approximation (Mom.-ind. app., gray band). The constraints from the HIC~\cite{Tsang2009_PRL102-122701}, the IAS~\cite{Danielewicz2014_NPA922-1}, and the ASY-EOS experiments~\cite{Russotto2016_PRC94-034608} are depicted with blue, yellow, and gray shadows, respectively.}
  \label{Fig2}
\end{figure}

The unique determination of the single-particle potentials in the full Dirac space gives us confidence to investigate the symmetry energy and its density dependence.
In Fig.~\ref{Fig2}, the red solid line shows the symmetry energy $E_{\text{sym}}(\rho)=\left.\frac{1}{2}\frac{\partial^2 E(\rho,\alpha)}{\partial\alpha^2}\right|_{\alpha=0}$ calculated by the RBHF theory in the full Dirac space as a function of the density $\rho$.
At the saturation density $\rho_0=0.188\ \text{fm}^{-3}$ (see Ref.~\cite{WANG-SB2021_PRC103-054319}), the symmetry energy $E_{\text{sym}} (\rho_0)$ is $33.1$ MeV, which is in agreement with the empirical values $31.7\pm3.2$\ MeV~\cite{Oertel2017_RMP89-015007}. The slope parameter of the symmetry energy $L$ is $65.2$ MeV, which is consistent with the empirical values $58.7\pm28.1$\ MeV~\cite{Oertel2017_RMP89-015007}.
Comparing to the results obtained by the projection method (olive dashed line), our results lead to a softer symmetry energy. This fact is also favored by the historical detection of the gravitational wave from GW170817~\cite{Abbott2017_PRL119-161101,Fattoyev2018_PRL120-172702}.
The gray band reveals the uncertainties of the momentum-independence approximation as discussed in~Ref.~\cite{WANG-SB2021_PRC103-054319}. 
These results again demonstrate the importance of the full Dirac space.

\begin{figure}[htbp]
  \centering
  \includegraphics[width=8.0cm]{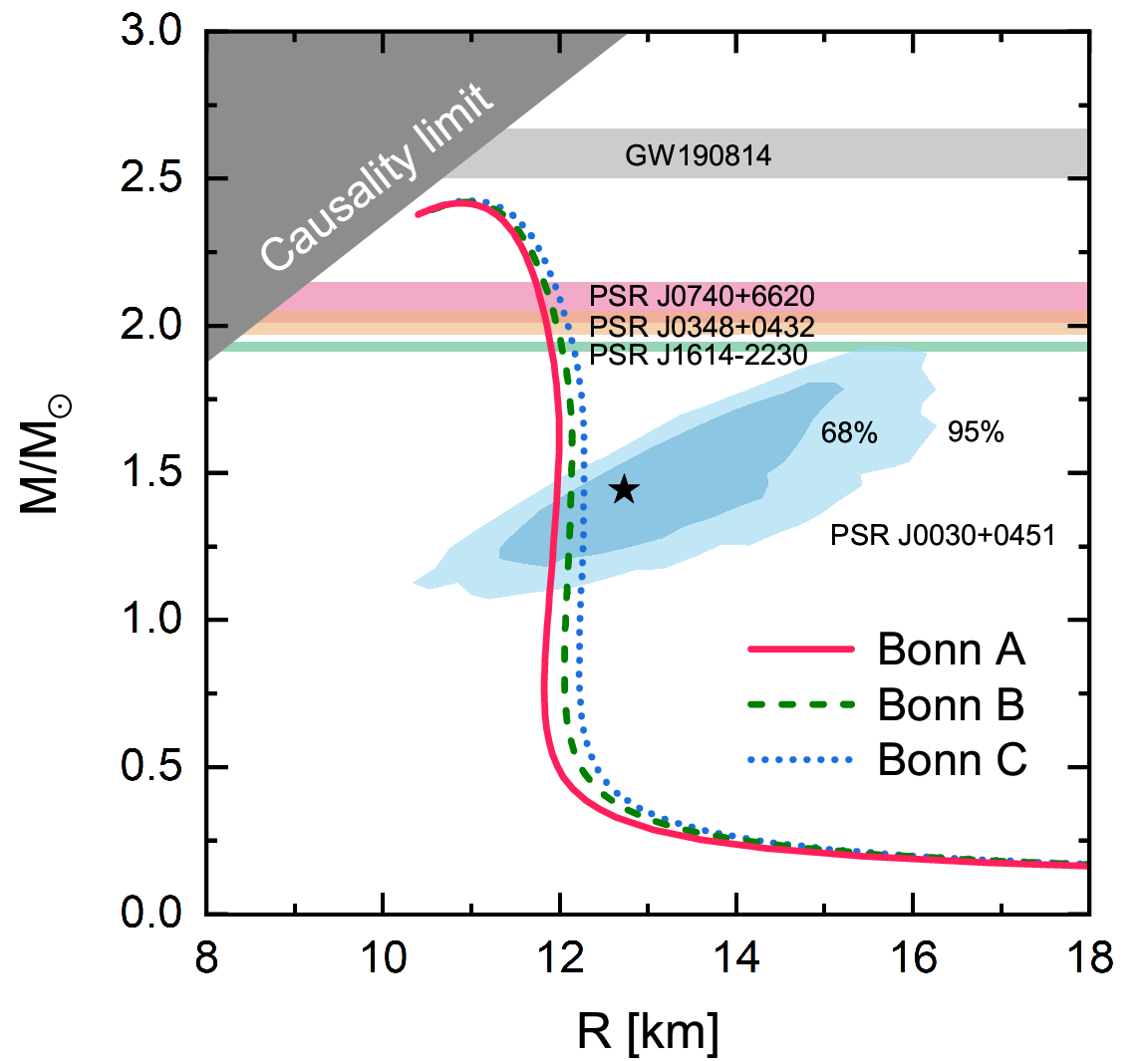}
    \caption{(Color online) The mass-radius relations of the neutron star obtained from the EOS by RBHF theory in the full Dirac space with the potentials Bonn A, B, and C. Astrophysical constraints from massive neutron star observations taken from Refs.~\cite{Demorest2010_Nature467-1081,Fonseca2016_ApJ832-167,Antoniadis2013_Science340-1233232,Cromartie2020-NatAstr4-72,Fonseca_2021-ApJL915.L12} are shown as horizontal shaded areas. The 
   dark and light blue regions indicate the 68\% and 95\% confidence intervals constrained by the NICER analysis of PSR J0030+0451~\cite{Miller2019-ApJ887-L24}. The recently inferred mass of the secondary compact object of GW190814~\cite{Abbott2020_ApJL896-L44} is also shown.
  The gray region is excluded by causality that $R>2.9\ GM/c^2$. See the text for details.}
  \label{Fig3}
\end{figure}

The symmetry energy has been extensively studied both from the theoretical and experimental points of view.
From the experimental side, we consider the data from simulations of the low-energy HIC involving $^{112}$Sn and $^{124}$Sn~\cite{Tsang2009_PRL102-122701},
nuclear structure studies involving excitation energies to isobaric analog states (IASs)~\cite{Danielewicz2014_NPA922-1},
and the ASY-EOS experiments at GSI~\cite{Russotto2016_PRC94-034608}.
They are shown as the blue, yellow, and black shadow regions in Fig.~\ref{Fig3}, respectively.
Below the saturation density, the symmetry energy obtained by the RBHF theory in the full Dirac space is found compatible with the constraints from the IAS~\cite{Danielewicz2014_NPA922-1} and the HIC~\cite{Tsang2009_PRL102-122701} experiments. At twice normal saturation density, i.e., $0.32\ \text{fm}^{-3}$, the symmetry energy obtained in this work is $51.6$ MeV, which is in agreement with the constraint $50.8$--$60.4$\ MeV from ASY-EOS~\cite{Russotto2016_PRC94-034608}.

The clear difference of symmetry energy for high densities shown in Fig.~\ref{Fig2} and its importance in neutron stars inspire us to apply the RBHF theory in the full Dirac space to neutron stars.
Based on the binding energy per nucleon as a function of the density and assuming $\beta$ equilibrium in the neutron star matter, the mass and radius of a cold, spherical, and static neutron star can be obtained by solving the Tolman-Oppenheimer-Volkov (TOV) equation~\cite{Oppenheimer1939_PR55-374,Tolman1939_PR55-364}.
In Fig.~\ref{Fig3} we show the neutron star mass-radius relations obtained from the EOS obtained by the RBHF theory in the full Dirac space with the potentials Bonn A, B, and C.
The radii of a $1.4M_{\odot}$\ neutron star $R_{1.4}$ are 11.97, 12.13, and 12.27 km, respectively.
Recently, the Neutron star Interior Composition Explorer (NICER) mission has reported two independent Bayesian parameter estimations of the mass and equatorial radius of the millisecond pulsar PSR J0030+0451 as $1.34_{-0.16}^{+0.15} M_\odot$ and $12.71_{-1.19}^{+1.14}$\ km~\cite{Riley_2019-ApJ887.L21}, as well as $1.44^{+0.15}_{-0.14}M_{\odot}$ and $13.02^{+1.24}_{-1.06}$\ km~\cite{Miller2019-ApJ887-L24}.
The confidence intervals for 68\% and 95\% about the relations between mass and radius from Ref.~\cite{Miller2019-ApJ887-L24} are also shown in Fig.~\ref{Fig3}.
It can be seen that the predictions from the RBHF theory are completely consistent with the constraints by NICER. 
Moreover, we notice that there exist many other estimates of $R_{1.4}$ from different sources (see Ref.~\cite{Al-Mamun_2021-PhysRevLett.126.061101} and references therein).
It is found that our results on $R_{1.4}$ in the full Dirac space are consistent with those works, except for a few with an upper limit smaller than 11.9 km.
The radii $R_{1.4}$ of a $1.4M_{\odot}$\ neutron star from the RBHF theory in the full Dirac space, the projection method, and the momentum-independence approximation are 11.97, 12.38, and 12.35 km, respectively. 
The relatively small neutron star radius suggested by the full Dirac space implies that the symmetry energy at higher densities is soft, which is consistent with the result shown in Fig.~\ref{Fig2}.

% Recently, the mass and radius of PSR J0030+0451 were observed simultaneously by the Neutron star Interior Composition Explorer (NICER) as $1.44^{+0.15}_{-0.14}M_{\odot}$ and $13.02^{+1.24}_{-1.06}$km~\cite{Miller2019-ApJ887-L24}. The confidence intervals for 68\% and 95\% about the relations between mass and radius from the NICER analysis are also shown in Fig.~\ref{Fig3}. It can be seen that the predictions from the RBHF theory are completely consistent with the constraints by NICER.

The core densities of massive neutron stars could reach 5--10 times the nuclear matter saturation density~\cite{Lattimer2004_Science304-536}, which is far away from the region where the Brueckner theory is applicable.
We follow the strategy proposed in Ref.~\cite{Rhoades1974_PRL32-324} and applied in Ref.~\cite{Gandolfi2012_PRC85-032801}, where the neutron star matter EOS above a critical density $\rho_c$ is replaced with the maximally stiff or causal one by $p(\epsilon) = c^2\epsilon - \epsilon_c$.
$p$ is the pressure, $\epsilon$ is the energy density, $c$ is the speed of light, and $\epsilon_c$ is a constant.
This EOS is maximally stiff and predicts the most rapid increase of pressure with energy density without violating the causality limit $c_s/c=\sqrt{\partial p/\partial \epsilon}\leq 1$ where $c_s$ is the speed of sound, which results in $R>2.9\ GM/c^2$ with $G$ the universal gravitational constant.
In our calculations, $\rho_c = 0.57\ \text{fm}^{-3}$ and the constant $\epsilon_c$ is a parameter and determined by ensuring that the energy density is continuous. These assumptions provide an upper bound on the maximum mass of the neutron star.

As shown in Fig.~\ref{Fig3}, the maximal neutron star masses $M_\text{max}$ obtained from the RBHF theory with the potentials Bonn A, B, and C are $2.43M_\odot$, $2.43M_\odot$, and $2.44M_\odot$, respectively.
These values are effectively reduced compared to the $3.2M_\odot$ obtained in Ref.~\cite{Rhoades1974_PRL32-324}, where the EOS of neutron star matter is assumed to be the one of free degenerate neutrons continued with the maximally stiff EOS for densities higher than $0.275\ \text{fm}^{-3}$.
Meanwhile, our results are consistent with the available astrophysical constraints from massive neutron star observations, such as PSR J1614-2230~\cite{Demorest2010_Nature467-1081,Fonseca2016_ApJ832-167}, PSR J0348+0432~\cite{Antoniadis2013_Science340-1233232}, and PSR J0740+6620~\cite{Cromartie2020-NatAstr4-72,Fonseca_2021-ApJL915.L12}.
Furthermore, from our calculation, the secondary compact component of GW190814~\cite{Abbott2020_ApJL896-L44} with the mass $2.50$--$2.67M_{\odot}$ might not be a neutron star.

%=======================================================================================
%\section{Summary}\label{SecV}
%=======================================================================================

\emph{Summary.} The relativistic Brueckner-Hartree-Fock theory in the full Dirac space is developed and applied to investigate asymmetric nuclear matter and neutron star properties for the first time.
The isospin dependence of the single-particle potentials is uniquely determined and the neutron-proton Dirac mass splitting $M^*_{D,n}<M^*_{D,p}$ is obtained.
The controversy between the projection method and the momentum-independence approximation concerning the isospin splitting of Dirac mass in asymmetric nuclear matter has been clarified. 
% The momentum dependence of the single-particle potential and the spacelike component of the vector potential are important to get the correct sign.
The symmetry energy $E_{\text{sym}}$ and its slope parameter $L$ at saturation density are $33.1$ and $65.2$ MeV, respectively, both in agreement with the empirical values.
Below saturation density, the symmetry energy is consistent with the experimental constraint of nuclear structure and heavy-ion collisions.
The mass-radius relations from the RBHF theory are consistent with the astrophysical observations.
Especially, the radius of a $1.4M_\odot$ neutron star is predicted close to $12\ \text{km}$, and the upper bound of the maximum mass of the neutron star is found to be less than $2.4M_\odot$.
In this case the secondary compact component of GW190814 with the mass $2.50$--$2.67M_{\odot}$ might not be a neutron star.

\begin{acknowledgments}

This work was supported in part by the National Key R\&D Program of China under Contracts No. 2017YFE0116700 and No. 2018YFA0404400, the National Natural Science Foundation of China (NSFC) under Grants No. 11935003, No. 11975031, No. 11875075, No. 12070131001, No. 12047564, and No. 12147102, the Fundamental Research Funds for the Central Universities under Grants No. 2020CDJQY-Z003 and No. 2021CDJZYJH-003, the MOST-RIKEN Joint Project "Ab initio investigation in nuclear physics," 
the Deutsche Forschungsgemeinschaft (DFG, German Research Foundation) under Germany’s Excellence Strategy EXC-2094-390783311, ORIGINS, and the Institute for Basic Science (Grant No. IBS-R031-D1).
Part of this work was achieved by using the High-performance Computing Platform of Peking University, and the supercomputer OCTOPUS at the Cybermedia Center, Osaka University under the support of Research Center for Nuclear Physics of Osaka University.

\end{acknowledgments}

\bibliography{Bib-ANM}

% \end{CJK*}
\end{document}